\documentclass[11pt]{article} 

\usepackage[utf8]{inputenc} 
\usepackage[T1]{fontenc}
\usepackage{currvita}
\usepackage[noblocks]{authblk}     

\usepackage{graphicx}
\usepackage{caption}
\usepackage{subcaption}

\usepackage{geometry} 
\usepackage{booktabs} 
\usepackage{array} 
\usepackage{paralist} 
\usepackage{verbatim} 
\usepackage{subfig} 
\usepackage{url}

\usepackage{fancyhdr} 
\usepackage{sectsty}
\allsectionsfont{\sffamily\mdseries\upshape} 

\usepackage[nottoc,notlof,notlot]{tocbibind} 
\usepackage[titles,subfigure]{tocloft} 

 \title{Many-to-one contagion of economic growth rate across trade credit network of firms}
\author{Nata\v{s}a Golo}
\affil{Cevipof center for political research, SciencesPo, Paris, France} 
\author{Guy Kelman}
\affil{The Hebrew University Jerusalem, Israel}
\author{David S. Br\'ee}
\affil{Dept. of Computer Science, University of Manchester, Manchester, UK}
\author{Leanne Usher}
\affil{Queens College, City University of New York, US}
\author{Marco Lamieri}
\affil{Dept. of Research, Intesa SanPaolo, Milano, Italy}
\author{Sorin Solomon}
\affil{Racah Institute for Physics, Hebrew University Jerusalem, Israel}

\begin{document}
 \maketitle

\section*{Abstract}
We propose a novel approach and an empirical procedure to test direct contagion of growth rate in a trade credit network of firms. Our hypotheses are that the use of trade credit contributes to contagion (from many customers to a single supplier - ``many to one'' contagion) and amplification (through their interaction with the macrocopic variables, such as interest rate) of growth rate. In this paper we test the contagion hypothesis, measuring empirically the mesoscopic ``many-to-one'' effect. The effect of amplification has been dealt with in our paper \cite{Golo:2014ys}.

Our empirical analysis is based on the delayed payments between trading partners across many different industrial sectors, intermediated by a large Italian bank during the year 2007. The data is used to create a weighted and directed trade credit network. Assuming that the linkages are static, we look at the dynamics of the nodes/firms. On the ratio of the 2007 trade credit in Sales and Purchases items on the profit and loss statements, we estimate the trade credit in 2006 and 2008.

Applying the ``many to one'' approach we compare such predicted growth of trade (demand) aggregated per supplier, and compare it with the real growth of Sales of the supplier. We analyze the correlation of these two growth rates over two yearly periods, 2007/2006 and 2008/2007, and in this way we test our contagion hypotheses. We could not find strong correlations between the predicted and the actual growth rates. We provide an evidence of contagion only in restricted sub-groups of our network, and not in the whole network. We do find a strong macroscopic effect of the crisis, indicated by a coincident negative drift in the growth of sales of nearly all the firms in our sample.

\section{Introduction}
In this paper we use the trade credit network of Italian firms to test a model of ``many-to-one'' contagion of economic growth or economic crisis. Academic research on inter-firm trade credit networks is still in its infancy, as the data on trade-credit transactions is not easily accessible. An exception is the analysis of the trade credit network of Japanese firms, which have been rather extensively characterized by, for example, Tamura et al. \cite{Tamura:2012ys}, Miura et al. \cite{Miura:2012kx} and Watanabe et al. \cite{Watanabe:2012fk}.

Our approach to direct contagion of economic growth rate is built on the assumptions based on the previous results published in various economic literature on balance sheet contagion (e.g. Kiyotaki \& Moore \cite{Kiyotaki:1997km}, Boissay \cite{Boissay:2006ix}, Petersen \& Rajan \cite{Petersen:1997kx}, economic growth Schumpeter \cite{Schumpeter:1942bh}), and from our own previously developed methods and models following complex systems approach (Solomon \& Richmond \cite{Solomon:2002rp}, Challet et al. \cite{Challet:2009pd}).

When many firms simultaneously borrow from and lend to each other, and in particular when these firms are speculative and dependent on the credit flow, shocks to the liquidity of some firms may cause the other firms to also get into financial difficulties. As obvious as this argument might sound, it is very difficult to prove the effect of direct contagion in the data, exactly for the reason that the linearity is lost as soon as the firms are simultaneously interacting with each other.
 
  The ``many-to-one'' contagion relies on the hypothesis that the  change in the annual sales of a supplier follows the change in
  the {\bf mesoscopically aggregated demand}, i.e. that the yearly growth of sales of a supplier would be proportional to the yearly growth of demand of its customers.  Assuming that the growth of purchases of all customers of a supplier in a yearly period is known to us, shouldn't we be able to predict the growth in the sales of the supplier? Indeed we should, as long as:
  \begin{itemize}
  \item  the linkages between the customers and the supplier are constant over the longer period (at least two years, as the growth rate of financial indicators reported in balance or profit and loss statements is available on yearly basis)
  \item the growth  of demand (purchases) of a customer is assumed to be uniformly distributed among all its suppliers. 
  \end{itemize}
  
  Assuming that the above conditions are satisfied, we compare the prediction of the growth of sales with the real
  growth as measured from the profit and loss statements, and so we can test our hypotheses of growth contagion from customers to suppliers.
 
  The paper is organized as follows. In section \ref{sec:theory} we
  describe the nature and applications of different interaction
  mechanisms that can be involved in the self-amplifying auto-catalytic
  loops in supporting both peer interactions and the bi-directional
  feedback between the micro and the macro structures of the economy.
  Following that, we describe the possible methodologies that could
  be used in order to empirically show the existence and operation
  of an auto-catalytic feedback. In section \ref{sec:data}, we give the detailed description
  of the data that were used for the network model and internal
  properties of the nodes. In that section the reader can also find
  the definitions of the variables and the algebraic notations. In section \ref{sec:results}, 
 we elaborate on the empirical results and provide their
  interpretation and implications. Finally we discuss the results in section \ref{sec:discussion}.

\section{Applicability of the methods of statistical physics to stochastic processes in economics}
\label{sec:theory}
One of the major issues in economics is to understand how relativelly small and temporary endogeneous changes in technology or wealth distribution may generate macroscopic effects in aggregate productivity, asset prices etc. For this purpose, it is necessary to identify self-amplifying mechanisms, filtering out the main mechanisms that may trigger dynamics leading to a systemic change, from other interactions destined to drown in the noise of local, short lived perturbations.
This implies that the vast majority of microeconomic interactions that may affect the macroeconomic/systemic level do so by a kind of auto-catalytic positive feedback loop.
This idea was already proposed in various contexts, but was often dismissed in the absence of concrete mechanisms that realize it.
 
Below, we list three mechanisms that might effectively amplify microscopic events to macroscopic dynamics in economic systems. In this paper we deal only with the first mechanism, with the second one we dealt in \cite{Golo:2014ys}, and the third mechanism we might tackle in the future.
 
  \paragraph{`Peer-to-peer', `One to many' and `Many-to-one' interactions
  between firms in the network}

  The domino effect, or contagion can be best understood using powerful
  mathematical and statistical-mechanics tools developed in percolation
  theory. They allow the rendition of precise predictions that
  correspond to real world stylized facts: macroscopic transitions
  caused by minute parameter changes, fractal spatial and temporal
  propagation patterns, delays in growth or crisis diffusion between
  economic sectors or geographical regions.  For a formal discussion
  of social and market percolation models the reader is referred
  to Goldenberg et al. \cite{Goldenberg:2000uq}.  
  
  In this paper
  we deal with the many-to-one contagion principle (in
  the cases when a supplier has a single customer it reduces to
  peer-to-peer principle).   
  The peers are tied financially,
  and also physically by the goods they pass. We introduce the assumption that the peers are correlated though their growth rate and we empirically test it. We attempt and succeed to find only partial evidence in our data  that the Many-to-one is responsible for the propagation of the growth rate on the trade network consisting  of suppliers and their customers. This mechanism is only partly responsible for the congation of growth rate, and other mechanisms, such as the following ones, should also be considered.
   
  \paragraph{Inter scale macro-to-micro reaction between firms
  and the system}
  For a long while, one of the drawbacks of the dynamical models
  inspired by physics was the absence of 
  interaction between scales. A model that introduces this interaction
  was introduced by Solomon and Golo \cite{Solomon:kx}. This is a
  financial model extending the ideas of Minsky \cite{Minsky:1982mz},
  where not only the fate of individual firms (e.g. failure)
  influences the system state (risk aversion leads to rise in
  corporate interest rates), but also the state of the system is
  feeds back onto its own components (e.g. a rise in interest rate
  leads to more failures). Together with contagion across the
  network, the model 
  generates a bounty of predictions that agree with the stylized
  facts. For example: delaying or arresting the propagation of
  distress by targeted intervention in key individual components
  or in system properties (such as the interest rate). This model
  has been confronted with empirical data in \cite{Golo:2014ys} and
  provided a significant contribution in the interpretation of the
  economic collapse in 2008.
 
  \paragraph{Self-interaction: firms acting upon themselves}
  A model for self-dependent reproduction was proposed by Shnerb
  et al. \cite{Shnerb:2000fe}, generalizing on the ideas of Malthus
  that proliferation in the microscopic level in stochastic systems
  leads to the spontaneous emergence of a collection of adaptive
  objects.  The model is analytically tractable by statistical-mechanics
  method as mentioned above. It generates a host of qualitative
  (phase transitions) and quantitative predictions about the
  macroscopic behavior of the system. These predictions were precisely
  confirmed by empirical measurements in many cases: crossing
  exponentials between decaying and emerging economic sectors after
  a shock \cite{Louzoun:2007dq}, identity of the wealth inequality
  `Pareto-Zipf' exponent \cite{Klass:2007fk} and the market instability
  exponent \cite{Choi:2012kx} \cite{Huang:2001vn}, etc$\dots$. 
  The self-interaction mechanisms are not in the scope of
  this paper, and the microeconomics models of firms are not considered as well.

\subsection{The difficulty of the empirical confrontation}
  In the analysis of contagion, causality is the key. The mechanisms
  we sketched in the preceding subsection call for testing, but the
  analysis should account for causality, rather than correspondence.
 
    The microscopic behavior of agents in a network should be revealed in the structure of the 
  \textit{links} between the agents and their \textit{interaction} patterns. 
    In a financial network, the visible  communication between agents is through the invoices they issue and the payments they make.

The
structure of links between the nodes in a network is formally termed the `topology
  of the network'.
  The metrics (indicators) typically used for measuring topology are for example:
  node degree being the number of connections from/to each node, 
  clustering being the number of distinct groups of nodes, and connected-component sizes.
  However, the description of the topology is not the subject of our paper.
  We aim to justify the existence of the network and the application of the network analysis by trying to measure the significance of the interaction across the network nodes.
  In our work, we focus on a selection of certain
  neighborhoods that consist of buyers from a single supplier. Thus,
  the clustering coefficients and component size are used only for
  validating that the selection process does not destroy the 
  statistical properties of the complete sample.

  We are taking into account that the properties of firms influence the interaction between them. 
  In financial networks, the strength and the interaction patterns cannot be interpreted without understanding the 
 level of financial exposure to risk that the agents (firms) are in.
  The financial risk of providing and extending the trade credit relation is an interplay between a firm's own liquidity
  position, its social neighborhood (its industry and its particular buyers and suppliers), and systemic effects such as  the
  interest rate. 
  Since the risk exposure of a firm cannot be directly measured, the ability to meet obligations subject to the environment and
  the system we have used  a quantity available in our data termed the `RATING score'.
  This quantity will be defined in the data section below.

  \section{Description of the dataset}
  \label{sec:data}
  The data on individual firms  come from a dataset (further on abbreviated as BS) of Italian limited
  liability companies end-of-year balance sheet and Profit \& Loss
  statements, which is a part of a proprietary database \footnote{
  The data base is proprietary of one of the main Italian banks and
  all the analysis have been performed in a fully anonymous way
  fulfilling both the data license policy, the privacy constraints
  and the bank's research department policy. Only aggregated data
  has been disseminated and distributed to the research group}.
  The network is assembled from the sales invoices issued by suppliers to their customers when they sell an item.
  Some of these invoices were presented to a bank in order to acquire  trade-credit (TC). The borrower
  in most cases was the seller in the supplier-customer pair.
  It is these invoices that the bank recorded and which we have been able to analyse. 
  In this study we combined the two datasets (BS) and (TC) in order
  to select the suppliers that are most appropriate for this analysis.

  \paragraph{Balance sheets database}
  In Italy, all limited liability firms are obliged to submit their
  annual financial report (balance sheet) to the local Chamber of Commerce.
  Items contained in the firm's balance sheet are assets and liabilities
  of the firm such as: Equity, Net-Sales, Accounts Receivable, Inventory, Bank
  loans, Accounts Payable, Financial Costs, etc $\dots$.

  Other than balance data, these reports contain financial ratios.
  Financial ratios are ratios of quantities within the balance sheet
  items, such as Acid Test or the Receivables Conversion Period and
  their purpose is to help quickly estimate the financial status of the
  firm.

  The balance sheets are collected and stored 
  by an external agency. The firms, no matter whether defaulting
  or non-defaulting at the end of the period, are ranked with a
  `RATING score' ranging from 1 to 9 in increasing order of default
  probability: 1 is attributed to firms that are predicted to be highly solvent, and 9
  identifies firms displaying a serious risk of default. Notice
  that the ranking is an ordinal: firms rated as 9 are not
  implied to have 9 times the probability of defaulting as compared
  to firms rated 1. A good description of the RATING score
  is given in a paper by Bottazzi et al \cite{Bottazzi:2011la}.

  \par
  For the purpose of our analysis, and in accordance to the
  practitioners’ behavior, we divide all companies into three groups,
  using the Rating score: one group with an easy Access to Bank
  Credit (rating 1-3), one with the Access to Bank Credit (ABC) at
  medium risk (rating 4-6) and the third group at high risk and
  little or no Access to Bank Credit (rating 7-9).
  
  \begin{center}
  \begin{tabular} { l | l }
    Rating 1 - 3 & ABC $\rightarrow$ A \\
    Rating 4 - 6 & ABC $\rightarrow$ B \\
    Rating 7 - 9 & ABC $\rightarrow$ C
  \end{tabular}
  \end{center}

  \paragraph{Trade Credit data}
  The Trade Credit database\footnote{%
  TC data is private. All the analyses were performed within
  the bank on completely anonymous data and only aggregated information
  was disseminated to the research group.
  } (TC) contains all inter-firm delayed payment transactions during
  2007 that were intermediated by a large Italian bank. This bank
  covers about 15\% of the entire trade credit in Italy, according
  to the official statistics of the Bank of Italy.

  \paragraph{Algebraic notation}
  We use the symbols $i$ and $j$ to
  mark a supplier and a customer, respectively.
  Transactions between a supplier and a customer will be presented by magnitude and 
  direction of the money flow. A general event of goods or services
  sold to a customer can be graphically described by $j \rightarrow
  i$.

  The 2007 TC (Trade Credit) records are invoices that account for
  goods/services supplied by firm $i$ to firm $j$ that were presented to the bank for discounting in 2007. These transactions
  were booked as accounts receivable in firm $i$'s assets,
  and as accounts payable in firm $j$'s liabilities. We can write
  \begin{description}
    \item $R_{ji} \equiv R_{(j\rightarrow i)}$ the sum of all payable trade
    invoices and cash payments from a customer $j$ to a supplier
    $i$ accounting for purchases of 
    goods or services in 2007.
    \item $Pi = \sum_j R_{ji}$ the total invoices used by firm $i$
    as collateral on a credit line in 2007.
  \end{description}
  We may also define the following notation from the BS (Balance
  Sheet) records of firm $i$:
  \begin{description}
    \item $R_{iy}$: balance sheet item of total sales (cash and credit) of firm $i$ in year y
    \item $P_{iy}$: balance sheet item of total purchases (cash and credit) of firm $i$ in year y
  \end{description}

  \paragraph{Testing for completeness of firm-level information}
  To ensure that we have sufficient coverage in the TC database of
  each firm's sales, we selected the firms $i$ for which the expression
  \begin{equation}
   Matching_i = \frac{Pi}{R_{i,2007}}
   \label{eg:MatchingDef}
  \end{equation}
  is greater than a pre-defined value. We call this the `Matching
  threshold' or 
  `Matching'. We vary the threshold between 0 and 1 in order
  to create the best sample for our empirical analysis. We choose
  a representative sample with high level of completeness
  by setting the matching threshold so it is the maximal value that
  still renders a single large connected component of size $n$, and
  that the consequent connectivity of firms spans from 1 to approximately the
  size of the component ($\max k_{in} \approx n$).

  The matching ratio reaches values greater than than 1 (more than 100\% completeness), perhaps
  due to misalignment of the time windows between the trade credit network
  and balance sheet data. This will be discussed below.
  The analysis in this paper is, therefore, restricted to the sample of suppliers that have a matching proportion of 0.8 and up to 1.2 (i.e. 80\% to 120\%, or formally $0.8 < Pi /R_{i,2007} <1.2$).
  The reason to prefer a range with a top-cap is that much
  larger proportion signals to us a mismatch: that the sample of
  invoices for the selected firm may be incorrect. Possibly due to an
  error in invoice registration. 

  We did not expect to retrieve a large network from the firms
  exhibiting high matching proportion. Several factors reduce the
  matching. The two most prominent ones are:
  \begin{itemize}
    \item trade credit data do not record changes in cash holdings. They do,
    however, record an accounts receivable on the supplier's side and
    an accounts payable on the customer side. When accounts receivables are cashed by the customer, a `sale' item will be booked to the supplier.
    Now, 
    outstanding invoices and payments made in 2007 may have been booked
    as accounts receivables in 2006. 
    We estimate a delay of 3 months on average for
    discounts and 9 months for payments (cf. the misalignment of the time frames).
    \item the Sales ($R_{i,2007}$) are all sales of the firm including 
    sales performed in other monetary channels. The TC database
    holds information on invoices only.
  \end{itemize}

\section{Results}
\label{sec:results}

  Out of $132,710$
  firms that were initially available both as creditors in the TC
  data (have incoming links) and in the BS data (have sales in $2006
  \dots 2008$), only 671 companies fulfill the Matching
  range $0.8 < Pi /R_{i,2007} <1.2$. Lowering this threshold 
  would grow the sample exponentially but at the cost of 
  unavoidably entering suppliers with lesser proportions of
  their total debt owed.
  In Table \ref{tab:netParams} we list some statistics of
  this network, assuming that the links are directional. Out of the
  total 671 suppliers, 190 companies are of reasonable size
  in accordance with bank practices ($R_{i,2007} > 10^6$ EUR).

  \begin{table}[h]
    \centering
      \begin{tabular}{|r|c|}
      \hline
      Feature & Count \\
      \hline
      Number of supplier/creditor nodes & 671 \\
      \hline
      Number of customer/debtor nodes   & 10,762 \\
      \hline
      Number of links   & 12,198 \\
      \hline
      Multi-edge node pairs     & 150 \\
      \hline
      Average number of IN-neighbors   & 18.17 \\
      \hline
      Average number of OUT-neighbors & 1.2 \\
      \hline
    \end{tabular}
    \caption{\small Some basic parameters of the subset network.
    Multi-edge node pairs give the number of node-pairs that link
    in both directions (reciprocal). Comparing the number of
    customers to the number of links presents a picture of many
    tree-like subgraphs connected to each other with minimal number
    of loops.}
    \label{tab:netParams}
  \end{table}

\subsection{In-degree distribution}
\begin{figure}
  \begin{subfigure}[b]{0.5\textwidth}
    \includegraphics[scale=3.0]{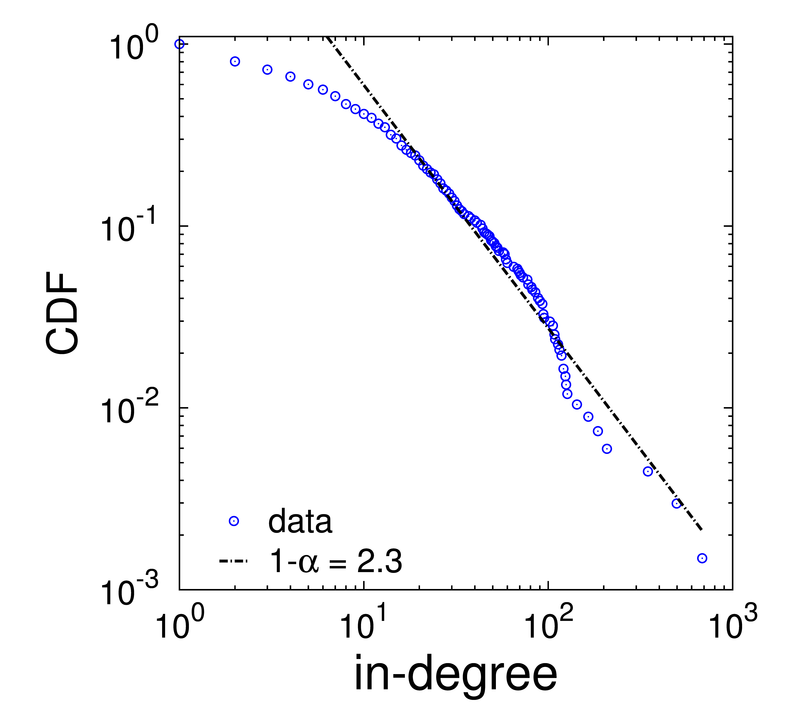}
    \caption{Cumulative histogram}
    \label{subfig:Zipf}
    \end{subfigure}
    \begin{subfigure}[b]{0.5\textwidth}
   \includegraphics[scale=0.35]{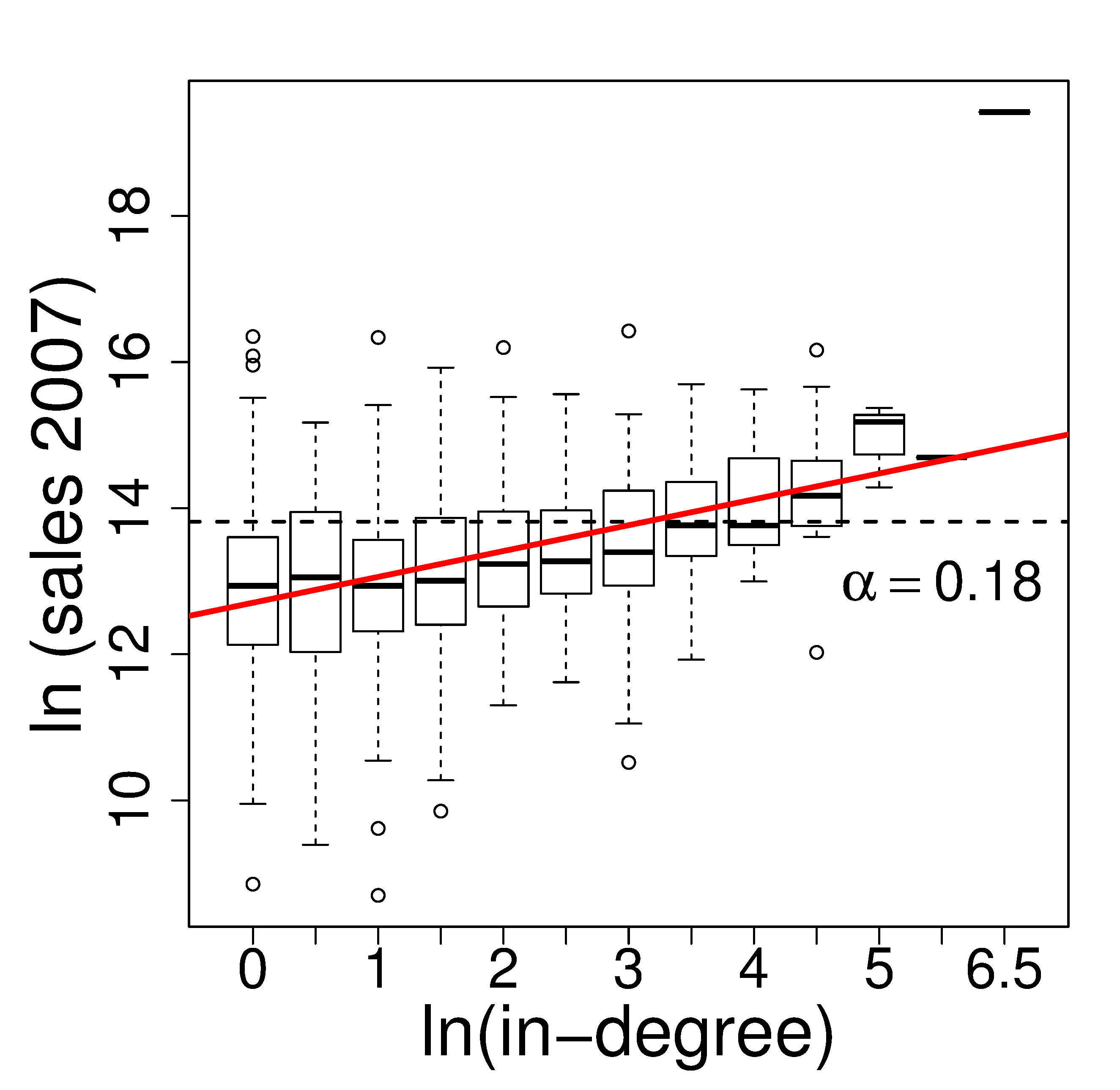}
   \caption{Size vs. in-degree}
    \label{subfig:INdegreeVSsales}
    \end{subfigure}
    \caption{\small Cumulative in-degree distribution of the 671
    suppliers adhering to the 80-120\% Matching criterion in the
    left panel. On right, a box plot of the same
    suppliers showing a correlation between their size distribution
    binned by in-degree (red curve represents the linear fit leaving out the degrees greater than $\exp(5) = 150$) }
    \label{fig:INdegree}
  \end{figure}

  In our inter-firm network, the in-degree of a supplier is the
  number of its customers. In a previous study, Miura et al.
  \cite{Miura:2012kx} obtained power laws of the in- and out-degree
  distributions, with exponents of $1.3$.  In
  our network, we were able to confirm a similar finding of $1-\alpha
  = 2.3$, i.e a negative exponent $\alpha = -1.3$.  As for the
  in-degree vs. size, a statistically significant correlation was
  recovered in our sample with a slope of $\exp(0.18) = 1.19$. 
  In the Japanese network of Miura et al., a slope
  of $1.3$ was obtained.
   Both results appear in figure \ref{fig:INdegree}.

  In the second observation, the correlation between
  Sales in 2007 and in-degree (panel \ref{subfig:INdegreeVSsales})
  Pearson's correlation coefficient is $r=0.32$.
  The broken line marks a net-sales of 1 million EUR. The crossing 
  between the broken and the red linear fitting line is characteristic
  of a firm with $K_{in} \approx 20$.
  Below the dashed line we can find the small companies, and it is
  clear that most of them also appear left of the crossing point,
  indicating a client-base smaller than $20$.

  \begin{figure}[h]
    \centering
    \includegraphics[scale=0.40]{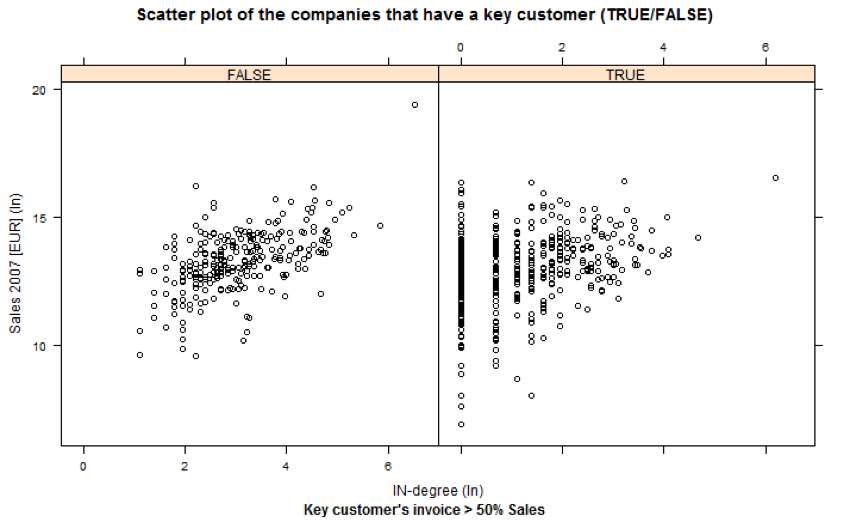}
    \caption{\small Log in-degree vs. log size of the sellers
    (creditors) that don't (FALSE) or do (TRUE) have a key-customer
    (FALSE on the left panel, TRUE on right).}
    \label{fig:TRUEFALSEkeyCustimers}
  \end{figure}

  \subsection{Key customers}
 In order to quantify the relevance of the many-to-one approach, we measure the number of suppliers in our selection that do not have a key customer but are permanently related with a large number of customers.
A supplier with a single key-customer is a firm that is to a large extent (50 \%) dependent
  on a single customer. Such a seller could be more fragile towards
  his customer's financial environment than a supplier who has no single key customer. 
 
However, the nature of the key-customer relations is not known to us.  It could be the nature of small businesses, though it is obvious from figurel \ref{subfig:INdegreeVSsales} that the firms that have low IN-degree correspond to a very large confidence interval (meaning that there are firms that have very large sales, in the order of million euro, and yet a single customer). This might have a significant (negative) impact on the experiments that we are performing. Namely, it might indicate that that there are some incidental events of sales that do not reflect the regular trading pattern. The vast number of key-customers in our dataset could also be connected with the fact that the data are preceding the economic crisis and it is probable that an increased number of liquidations, mergers etc happened which can be reflected in the incidental transactions.

  Figure \ref{fig:TRUEFALSEkeyCustimers} displays two subgroups of the suppliers presented
  in Figure
  \ref{fig:INdegree}. On the right panel
  are the suppliers that have a `key-customer', and on the left
  appear the suppliers that do not have such a customer.  We define
  a key-customer as one that  purchases of at least 50\% of
  the supplier's annual sales.
  Out of the 671 firms in the sample there are 414 creditors with
  a key-customer and 235 without one. Having a key-customer is a
  feature of the supplier and clearly all suppliers that have one
  customer ($K_{in} = 1$) qualify for it.
  We can appreciate that the range of firms that have a key-customer
  is dominated by suppliers with a small in-degree. However, we
  also find that suppliers with a large in-degree ($K_{in}>50$)
  have a key-customer.
  The reason is that in many situations, the payment distribution
  to a single supplier is fat-tailed, i.e. the largest customer
  pays an order of magnitude more than the second largest one.
  A good example of this is a phone company: it has few very large
  customers and the majority are single-time walk-in clients.
  We can then expect that the suppliers with a key-customer will be subject to transmission of financial signals
  from their key customer, either directly by peer-to-peer interaction
  or indirectly by responding quickly to the factors that influence
  the key-customer.

  In this small subgroup of the firms, as the
  ratio of suppliers with a key customer to those without is 2:1,
  we should  expect to see a contagion effect that spans
  the  firms with a key customer at the very least.

  \subsection{Contagion of growth or distress}
  We checked the contagion of the growth between the Customers and
  the Suppliers on the selected sample of suppliers with the Matching
  80-120\%. We compared the actual measured growth in net-sales (cash
  and credit) from 2007 to 2008 of each supplier $i$:
  \begin{equation}
    \ln \frac{R_{i,2008}}{R_{i,2007}}
    \label{eq:logGROWTH}
  \end{equation}
  against an estimated aggregated growth of purchases from all its customers,
  We assume that the change in purchases of a buyer $j$ between one
  year to the next is uniform across all his suppliers.
  So the estimated  purchases of buyer $j$ from
  supplier $i$ in 2008 is the payment received in 2007, $R_{ji}$, 
  weighed by the trend in purchases of $j$:
    \begin{equation}
    \nonumber
     \frac{P_{j,2008}}{P_{j,2007}}R_{ji}.
  \end{equation}

\begin{figure}
    \centering
    \begin{subfigure}[b]{0.5\textwidth}
    \includegraphics[scale=0.6]{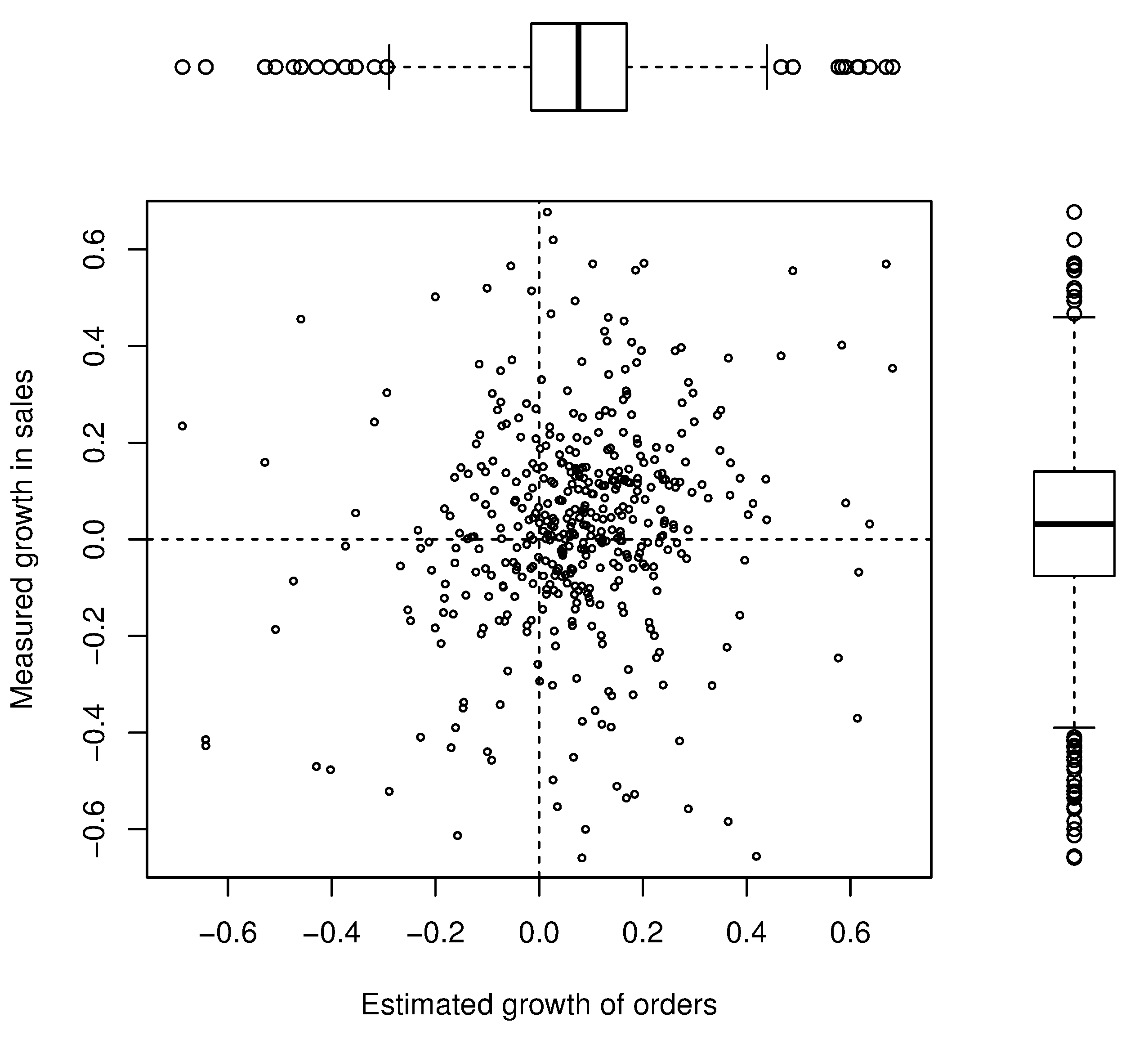}
        \caption{2006-2007}
    \label{subfig:3agrowth}
    \end{subfigure}
    \begin{subfigure}[b]{0.5\textwidth}
    \includegraphics[scale=0.6]{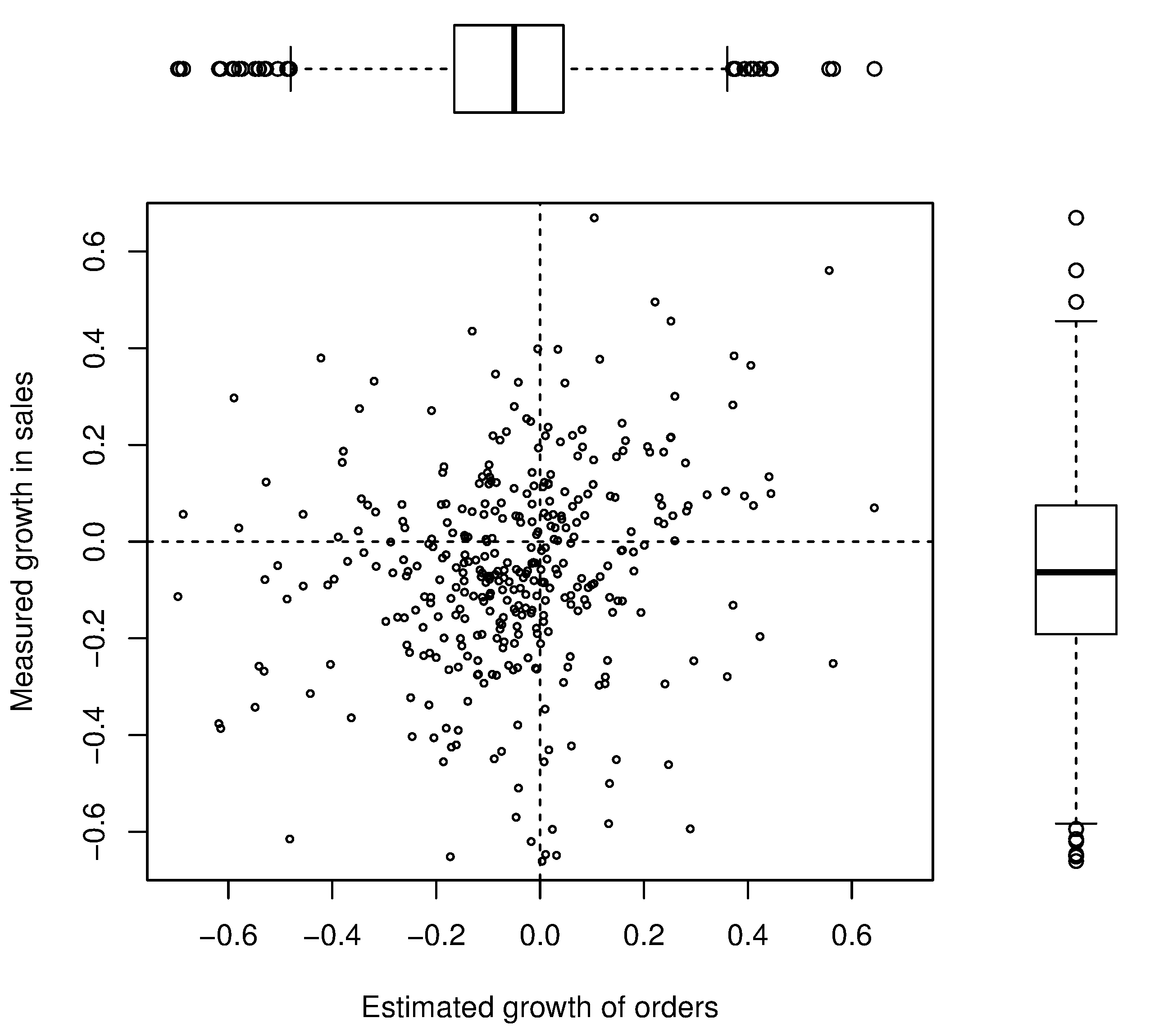}
        \caption{2007-2008}
    \label{subfig:3bgrowth}
    \end{subfigure}
    \caption{\small Growth rates in the Sales
    of supplier $i$, recorded in the financial statements, versus our estimation of growth of the collective / aggregated orders from all his customers. The side/top bar plots in both subfigures are used to estimate the centroid of the clouds of points. Figure \ref{subfig:3agrowth} has a centroid in the first quadrant, defined by a positive value of the mean growth rate of Sales, and a positive value of the mean growth rate of estimated demand. Figure \ref{subfig:3agrowth} has a centroid in the third quadrant, defined by a negative value of the mean growth rate of Sales, and a negative value of the mean growth rate of estimated demand.}
    \label{fig:3growth}
  \end{figure}

Summing over all supplier $i$'s customers gives an estimate of the growth in his sales from 2007 to 2008: 
  \begin{equation}
    \ln \frac{\hat{P}_{i,2008}}{P_i} = \ln \frac{\sum_j \frac{P_{j,2008}}{P_{j,2007}}R_{ji} } {\sum_j R_{ji}}.
    \label{eq:logPredictGROWTH}
  \end{equation}
  If we place (\ref{eq:logGROWTH}) on the Y-axis and
  (\ref{eq:logPredictGROWTH}) on the X-axis we obtain a
  2-dimensional scatter in which a dot will travel in the upward direction away from the origin to mark a firm with growing sales, and will travel to the right of the origin to designate growing \textit{predicted} growth in sales. This scatter plot is shown in figure \ref{subfig:3bgrowth}.

  Assuming that the links in the 2007 network are constant across the three-year frame (2006, 2007, 2008)
  and that customer's purchases correspond to supplier's sales,
  the resulting pattern is expected to form a straight line through
  the origin with a slope of $+1$.

We   applied the same aggregation and estimation procedure
  to the prior period. Applying similar reasoning, we write the
  growth in sales (2006-2007) as
  \begin{equation}
    \ln \frac{R_{i,2007}}{R_{i,2006}}
    \label{eq:logGROWTHm1}
  \end{equation}
  again, assuming that any change in the customer pool of a supplier
  between 2006 and 2007 is negligible, we write the aggregated
  growth in orders for that period as
  \begin{equation}
    \ln \frac{P_i}{\hat{P}_{i,2006}} = \ln \frac{\sum_j R_{ji}}{\sum_j \frac{P_{j,2006}}{P_{j,2007}}R_{ji} } 
    \label{eq:logPredictGROWTHm1}
  \end{equation}
  Figure \ref{subfig:3agrowth} displays this scatter plot.
  
  It is important to note the magnitude of the growth rates:
  for the majority of the suppliers, a small increase in orders to
  the supplier correspond with a small increase in the annual
  net-sales. This is the reason that the bulk of the points are
  close to the origin of the axes. i.e. the growth rate distribution
  is extremely narrow and deviations are dominated by the rare
  events. We should still expect that by the scaling nature of the
  growth rates \cite{arXiv:1004.5397v1}, the rare events will render
  the same occurrence as the frequent ones.
  
   In the subplots
    of figure \ref{fig:3growth} we added (top and right side of each subplot) a
  box-and-whisker plot on the sides to mark the univariate growth
  rate and predicted growth distributions.
  In these sidebars we can note two features: \textbf{(1)} that the
  estimated growth rate distribution is also narrow, i.e. shows
  similarity to the tent-shaped actual growth rate distributions, and
  that \textbf{(2)} the positions of the median growth rate values
  indicate that the centroid of the pattern in the period 2007/2006
  (panel \ref{subfig:3agrowth}), the median value of both, the estimated and the real growth of all suppliers is positive. The growth in the estimated sales is larger then the measured one. In contaxt to that, in the next period, 2008/2007 (panel \ref{subfig:GSyp1yVSYi}),
  the centroid sits in the third quadrant, i.e. the meadian values of the box plots in    \label{subfig:GSyp1yVSYi} show negative growth of both the estimated and the real sales of the suppliers.This is an evidence of the transition between an economic boom (a positive growth in sales and demand) in the first period and a bust (negative growth in demand and sales) in the second period.  
  The results show that a decline of growth in sales and purchases in the second period (an incline in the first) occurred concurrently for suppliers and buyers.
  \par
  This simultaneous switch in the typical behavior is the effect of the credit crunch: the
  purchasing power of the customers has shrunk as most of the suppliers were not able to allow credit
  to their customers.
 
  However, figure   \ref{fig:3growth} fails to prove the correlation between the estimated and the real sales of customers. Even by visual inspection, it
  is evident that the slope of $+1$ is not present. The conclusion we draw from the plot is that in the selected sample 
  little or no correlation exists between the aggregated changes
  in purchases by a supplier's customers and the growth rate in that
  supplier's Sales.

  Some companies, however,
  do not `follow the crowd' by going negative for several reasons apply, among which are  regulatory actions and the sectoral behavior which
  we will analyze in the sequel. 

 \subsection{Sectoral differences}
\begin{figure}[h]
    \centering
    \includegraphics[page=9,scale=0.50,bb=3 10 480 495]{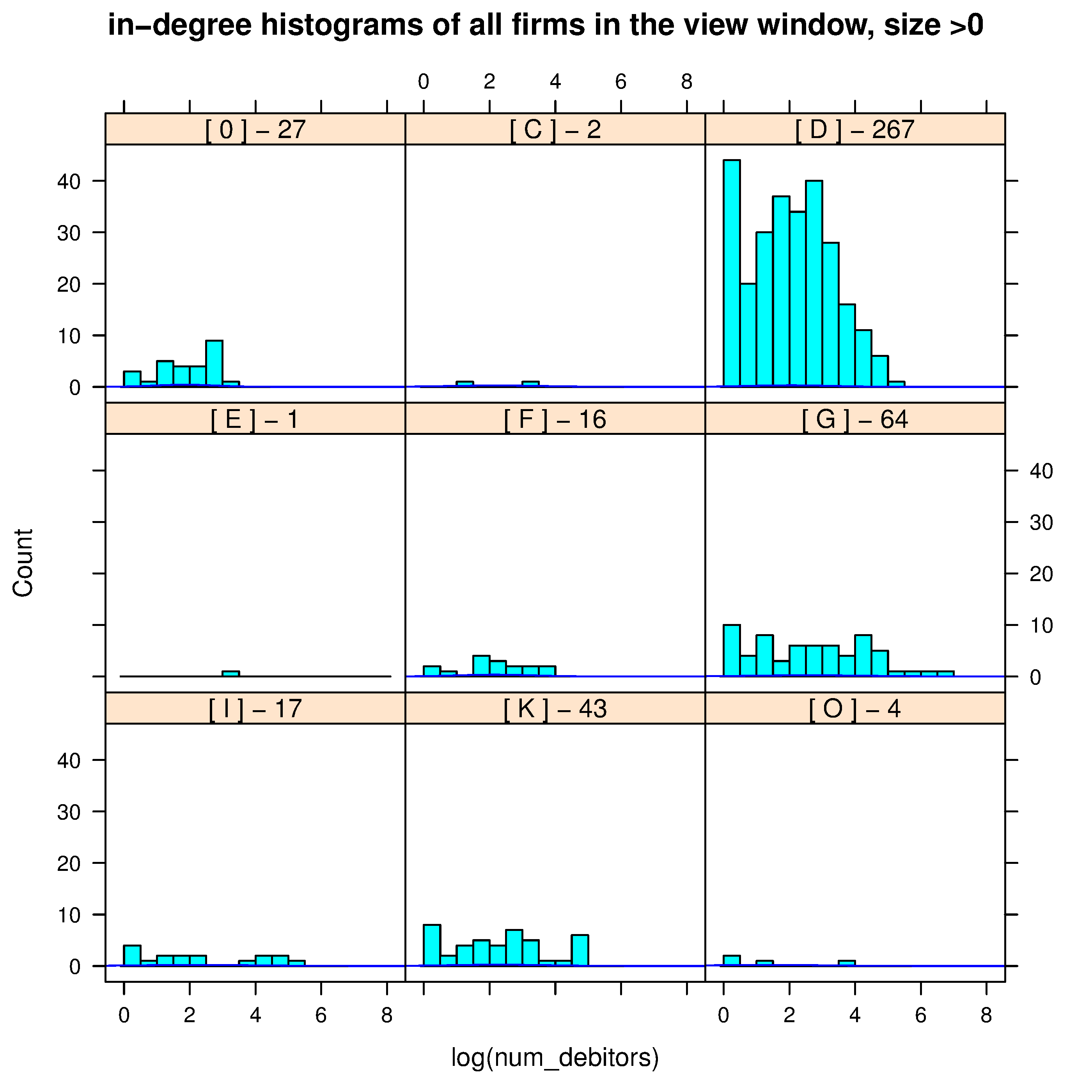}
    \caption{\small In-degree histograms of the subgroups according to their industrial classification: C - Mining and quarrying; D - Manufacturing; E - Electricity, gas and water supply; F - Construction; G - Wholesale and retail trade, repair of motor vehicles; H - Hotels and restaurants; I - Transport, storage and communication; K - Real estate, renting and business activities; O - Other community, social and personal service activities. The number of companies with in-degree$>0$ is given in the title of each panel.}
    \label{fig:indegreeHist9Panel}
  \end{figure}

  The sub-sample of the network
  ($0.8 < Pi /R_{i,2007} < 1.2$) consists of a heterogeneous set: there is a
  large variability in the connectivity pattern (in-degree)
  and in the net-sales. But the most relevant diversification factor (and possibly
  related to the previous ones) is that the companies come from
  different industries. About a half
  of the selected suppliers are in the Manufacturing sector (industrial
  classification numbers 15xx-37xx). The other half of the sample are in
  other industries, primarily in Construction, Wholesale, and Transport. The degree distributions and the composition of the sample satisfying the 80-120\%
  Matching criterion are given in figure \ref{fig:indegreeHist9Panel}.

  There is a striking difference in the connectivity between the sub-samples
  from different industrial sectors.
  The Manufacturing sector (D) is the most populated, and
  the in-degree histogram of the companies within this sample
  shows that the companies follow the general in-degree
  distribution of figure \ref{fig:INdegree}. The second largest sector is Wholesale trade and
  Retail (G). We note that the number of companies with
  large in-degree in sector G is exaggerated. Comparing with
  the histogram of sector D, in G, the number of firms with 
  $K_{in}>150$ is as large as that in D although the total number
  of firms is 4 times smaller. This is due to the different nature
  of their businesses as will be explained further in the text. The
  third largest sector is K. In our data set, most of the companies
  in this sector are IT (software) companies.
  Again, the accounting procedures in IT are different from the ones in
  Manufacturing. Software companies will often provide services
  rather than goods.

\begin{figure}[h]
    \centering
    \includegraphics[page=6,scale=0.50,bb=3 16 440 415]{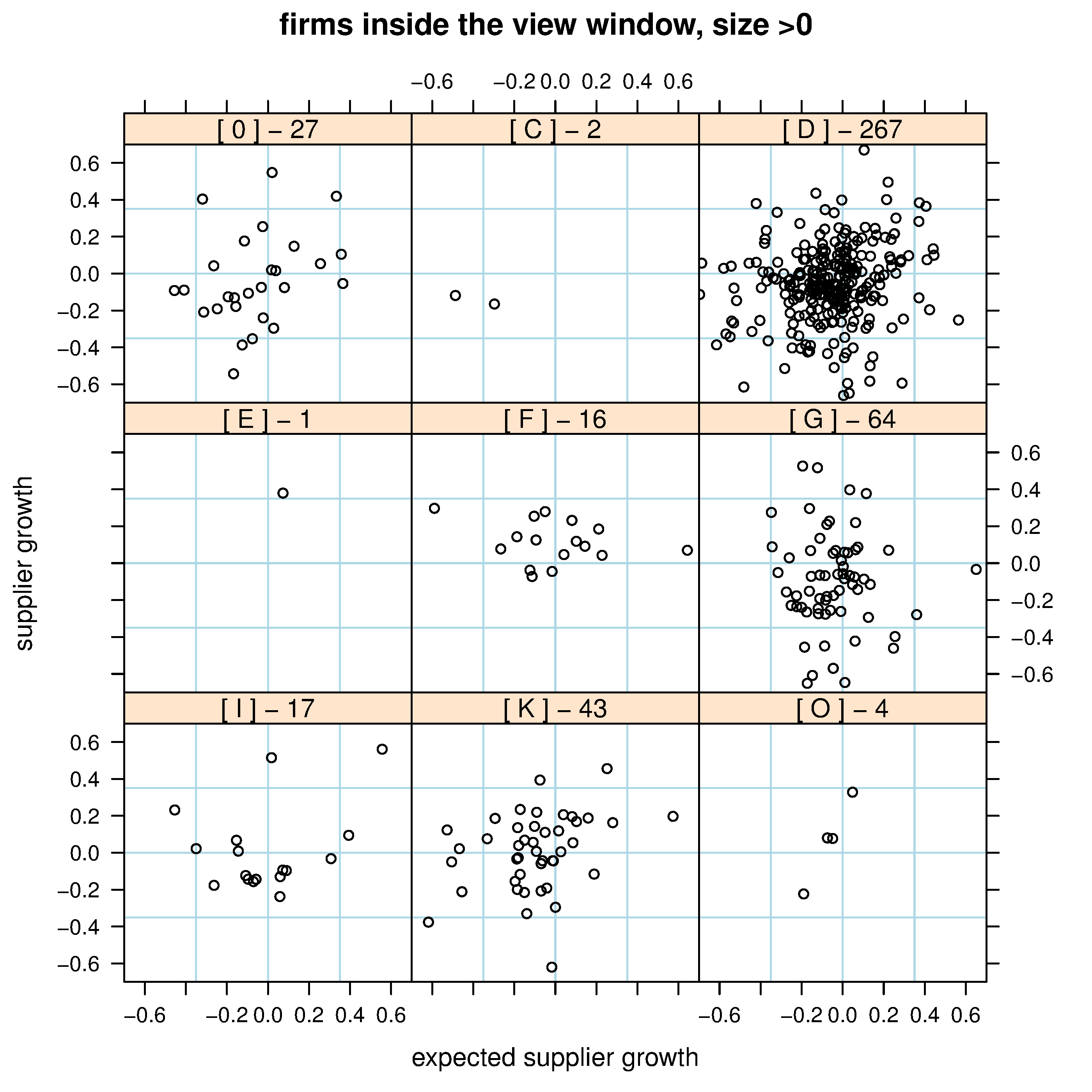}
    \caption{\small Growth between 2007 and 2008 in versus growth
    in Sales of supplier $i$ versus estimated growth of orders of all
    customers of a supplier $i$, per industrial sector. The sectors
    are the same as in Figure \ref{fig:indegreeHist9Panel}}
    \label{fig:firmsInView}
  \end{figure}

  In order to understand the differences in growth rate correlation
  between the suppliers coming from different industrial sectors
  and their customers, the scatter plot in figure \ref{subfig:GSyp1yVSYi}
  was split by sectoral subgroups. The subgroups are shown
  in figure \ref{fig:indegreeHist9Panel}, and the split growth rate
  scatter plots are given in figure \ref{fig:firmsInView}. In this figure,
  there are significant differences on the sectoral level.
  Most remarkable is the characteristic behavior of the G sector
  (the general name is `Wholesale and Retail', but in our database
  it is mostly composed of Wholesale firms). In this sector a
  significant number of firms has an estimated growth of orders
  close to one\footnote{shows as zero on a logarithmic scale, and the meaning is `no growth'}, but the measured growth in sales shows a notable variability
  away from no-growth.
  
  Another sector that shows uncommon behavior is sector F, 
  Construction. In this sector the growth response is opposite
  to the situation in sector G: there's a large variability in the expected growth of customer orders but the recorded
  growth of supplier Sales shows very little deviation from a state
  of no-growth.
  The atypical in-degree distribution in the Construction
  industry is discussed in Miura et al. \cite{Miura:2012kx}.
  In the Japanese industrial business network, they ran a flow algorithm
  and observed a difference between in-degree and the ability to
  source or receive money. In Construction, a large in-degree of a
  firm may not indicate large amounts of inflowing funds since at
  times they may relay funds via small single-customer subcontracting
  firms that dominate their payment distribution (firms that exist
  only to operate in a single one project).

  In our network we also observe greater intra-industry
  trade in Construction, more than would be expected by chance.
  However, as can be seen from figure \ref{fig:indegreeHist9Panel},
  sector F (center panel) is small in total number of firms.

  \begin{figure}
    \centering
    \begin{subfigure}[b]{0.5\textwidth}
    \includegraphics[scale=0.1]{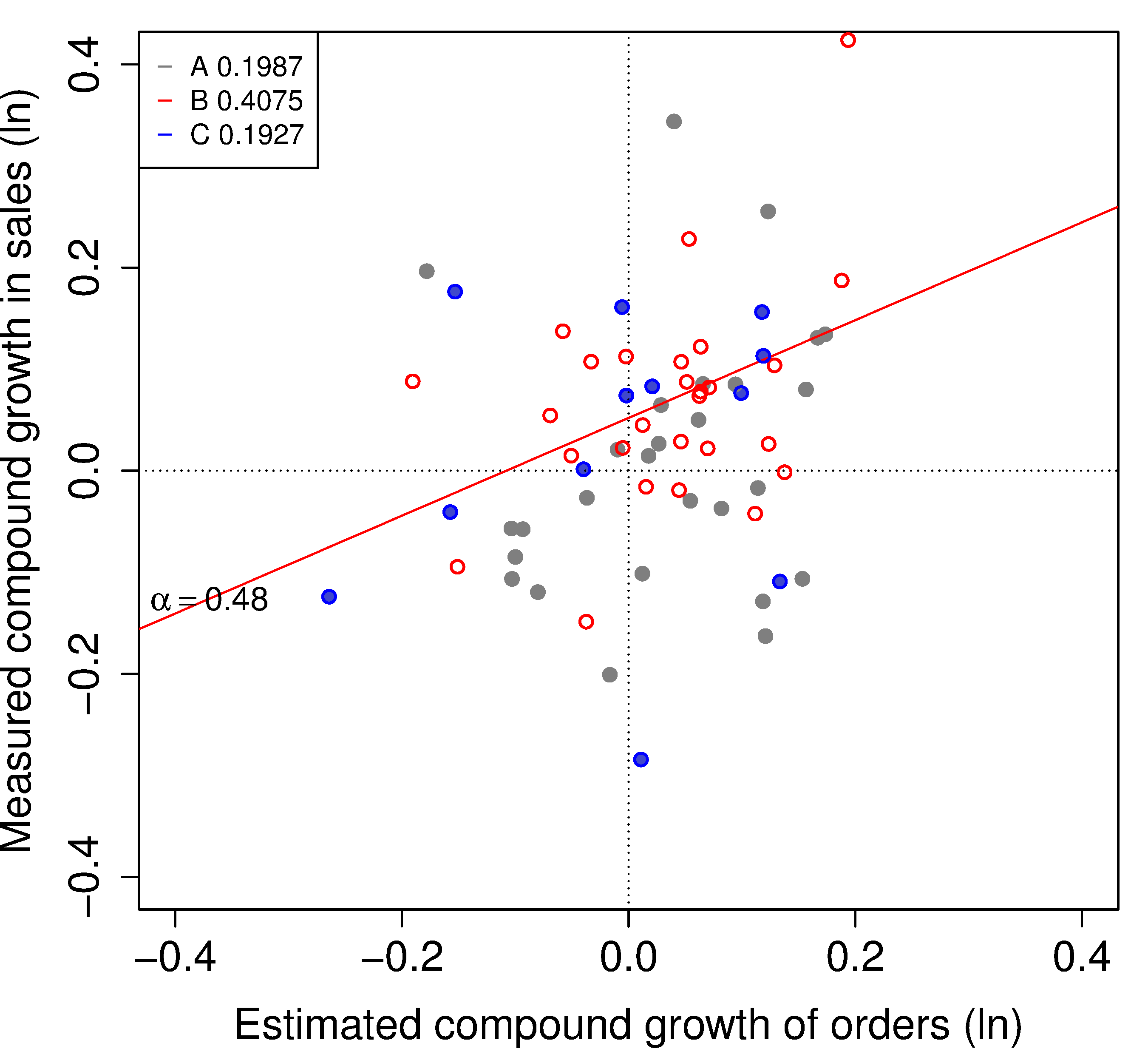}
    \caption{all firms}
    \label{subfig:6a}
    \end{subfigure}
    \begin{subfigure}[b]{0.5\textwidth}
    \includegraphics[scale=0.1]{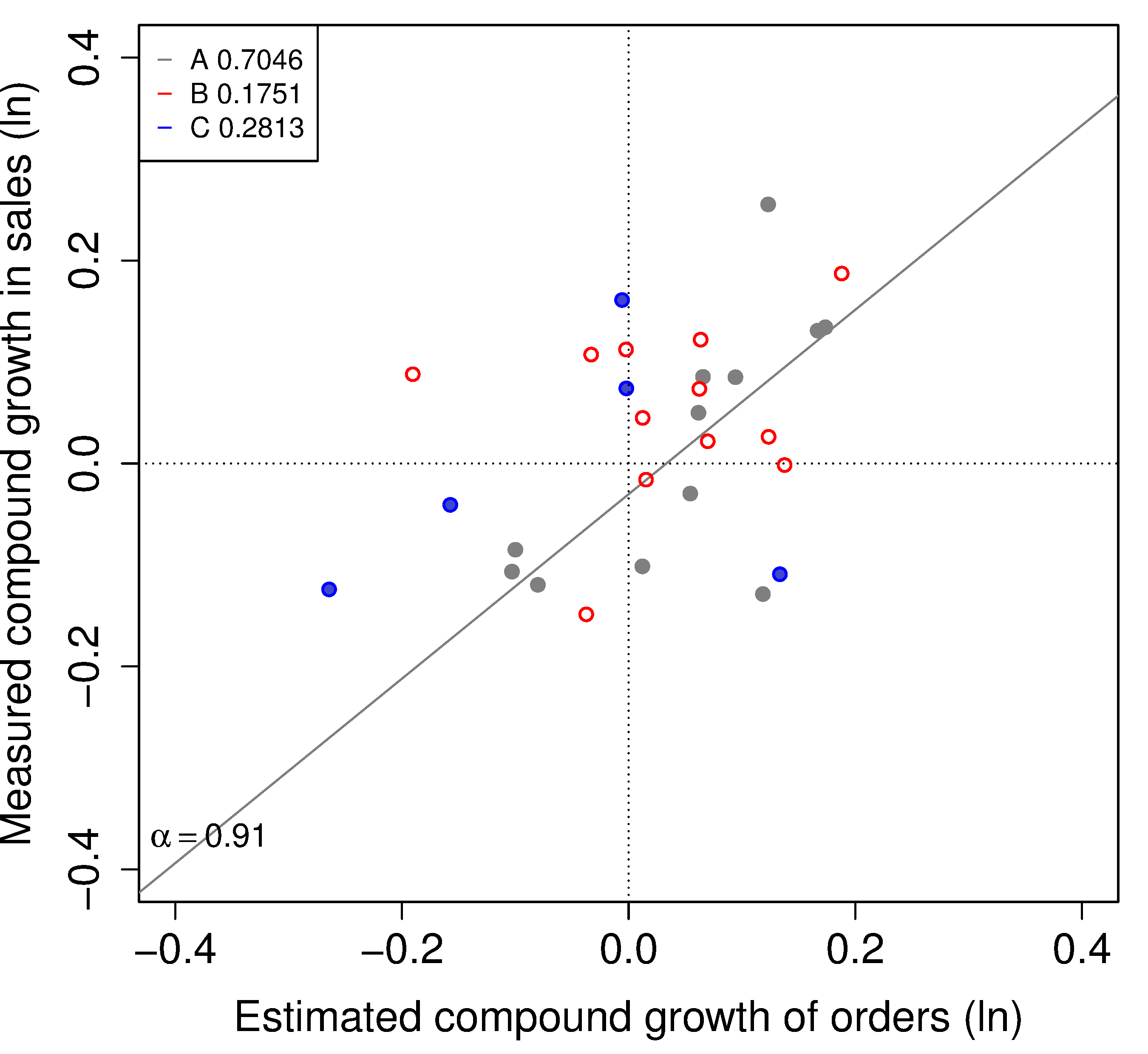}
    \caption{Sales $>10^6$ EUR}
    \label{subfig:6b}
    \end{subfigure}
    \caption{\small Measured compound growth in Sales of a supplier
    (y-axis) versus the estimated compound growth of the sum of the
    Purchases that his trade credit customers made to him (x-axis).
    Colors of the symbols correspond with the Rating of the
    companies in the following way: A - gray, filled circles, B - red, empty circles, C - blue, filled circles. The supplier companies are chosen to be only
    from the manufacturing sector (28xx) and their customers might be from any sector. 
    The right panel gives the companies with Sales$>10^6$ EUR. Correlation
    coefficients for each Rating/Size category appear in the legend in the upper left corner of each plot.}
    \label{fig:CAGR2Panel}
  \end{figure}

  In order to factor out the possible cause of heterogeneity in behavior 
  we chose suppliers that belong in a single industrial sector.
  Being the largest in the sample, we chose a sub-sector of
  Manufacturing, `Machinery and mechanical equipment'. This sector
  is well represented in the Italian trade network. We further tried
  to distill the effect of contagion by observing the geometric
  mean of growth rates in two consecutive periods 2007/6 and 2008/7. 
  This is commonly termed the Compound Annual Growth Rate (CAGR).  The plots of CAGR are given in figure \ref{fig:CAGR2Panel}.
  There we placed the CAGR of the supplier on the Y-axis versus the
  predicted CAGR from the sum of the purchases of his trade-credit
  customers on the X-axis.
  Colors of the dots mark the Rating class of the
  companies: ABC=A in gray, B in red, and C in blue.
  
  The scatter plot in figure \ref{fig:CAGR2Panel} may be interpreted
  in the following way: the companies in the \textit{first quadrant} have
  managed to keep positive growth of both sales and orders in the
  two-year period. In both panels we observe that points in this
  quadrant are red.  This means that the suppliers with Rating class
  B (4-6) managed to maintain
  their sales and orders with the same partners over the two year period and through the financial crisis. Although
  this result is in contrast with the hypotheses proposed in
  \cite{Shenoy:2011fk}, it did not come as a surprise.
  The homophily measurement on the same data set that is discussed
  in Kelman et al \cite{Kelman:2015fk} gives clear evidence that
  same Rating-class firms have a greater probability to attach to
  each other, with a small tendency of the customer to create business
  with a higher rated supplier.

  In the \textit{second quadrant}
  are locate suppliers that kept their growth in sales, opposing the downward trend in orders
  by their customers.
  Negative estimated growth but positive real growth could happen for many reasons. The most obvious one draws from our assumption that
  the network is static. This is only an approximation and although in
  the Manufacturing sector it is mostly a good one, in
  some cases it is possible that financial changes take place and
  trading partners change over time, especially the ones that are credit-constrained. 
  According to Delli Gatti et al.
  \cite{Gatti:2008yq}, the origin of fluctuations is due to the ever-changing
  configuration of the network of heterogeneous firms and the entire
  dynamics is being shaped by financial variables.
  The number of companies in this quadrant is also very small (only 4 firms with Sales larger
  than 1 million EUR out of only 13 firms in the sample).
  
  The suppliers in the \textit{third quadrant} were
  affected by the crisis the most. Both the orders from them and
  their own sales have decreased. In this range, surprisingly, we find the
  suppliers from the top Rating class `A'. Having a good credit-rating is
  related with the Current Ratio and liquidity of the firm as
  previously demonstrated by Beaver \cite{Beaver:1966kx} and later by Ohlson \cite{Ohlson:fk} and others. Also, having a good credit-rating corresponds
  with easy access to bank credit. i.e. the borrower is able to obtain a greater proportion of his collateral on loan, with a low interest. However, a
  good liquid
  position does not correspond with the state of the market in
  general, therefore these companies were not able to maintain
  their growth rate in the time frame of the crisis.

  Last, in the \textit{fourth quadrant} are the suppliers, which estimated
  aggregated orders grew while their sales decreased. 
  There could be several reason for this, though this might also be a sign that the network was not stable as expected, so the customers have purchased from a new supplier.

  The most interesting outcome of these measurements is the correlation
  of the CAGR with both Rating and firm size:
Observing the Pearson's correlation 
  coefficients, given in the keys of the subfigures of figure \ref{fig:CAGR2Panel}, we could see that the correlation
  between the compound annual growth rates in sales and orders are the highest
  ($\sim 0.70$) in the case of \underline{large} sized \underline{A-class} companies.
  In the second place come the medium credit-rated \underline{B-class}
  companies, in the case of all firms, with the correlation coefficient of $\sim 0.4$. The companies with Rating class C did not exhibit 
  correlations greater than $0.3$ either in the case of all firms or large firms. 
  
  An additional support for the very good correlation between the expected growth of customers' orders and the growth of sales in the companies with top Ranking (and therefore good access to credit), can be find in the literature. Recently, several scholars \cite{Shenoy:2011fk}, \cite{Garcia-Appendini:2012fj}
  attempted to model and empirically confirm the effect of the
  2007-2008 financial crisis on between-firm liquidity provision.
  Searching for a causal effect of a credit-rationing by the banks,
  Garcia-Appendini \& Montoriol-Garriga \cite{Garcia-Appendini:2012fj}
  tested the hypothesis that firms
  with high liquidity levels before the crisis, increased the trade credit
  extended to other corporations and subsequently experienced better
  performance compared to ex-ante cash-poor firms. They conclude that
  trade credit taken by constrained firms increased during this
  period. Therefore, they have intested in mainaining their customers and this might explain the high correlation which we measured.

\section{Conclusion}
\label{sec:discussion}

  This work examines correlation of the estimated growth rates of customers' orderd with the measured growth rate of suppliers' sales
  in a many-to-one setting,   where the customers of each supplier are
  responsible for at least 80\% of
  the supplier's sales, measured at the beginning of the second time period. By establishing their trade connections
  in during one year period, a growth rate prediction is made from a combination of the
  trend in purchases by these customers, and the social structure
  in the supplier's neighborhood. This prediction
  is compared with the
  real growth in sales of the suppliers. Results indicate
  the existence of growth rate contagion 
  only inside the following restricted
  sub-selections of the manufacturing firms, and compound over two years:
  \textbf{(1)} the large firms with A-class credit-rating, and \textbf{(2)} suppliers of any size that have 
  medium credit-rating (B-class).

  This gives evidence that direct contagion of growth
  rates between customers and suppliers is sensitive to the
  following factors:
  \begin{itemize}
    \item sectoral heterogeneity: in general, the industry of a
        supplier and a customer may or may not be the same. Each
        industrial sector has a behavior that is characteristic to
        it due to the accounting procedures required in that industry.
    \item in data coming from the bank, microscopic effects are
      still secondary compared to macroscopic effects, even during a
      state of crisis since the problematic customers may avoid
      approaching the bank. 
    \item missing data: while carefully considering numerical
      drift and retaining the overall statistics, the cleaned and
      filtered sample is still three orders of magnitude smaller
      than the number of firms in the full data set.
  \end{itemize}
  
  We are also aware that performing the mesoscopic many-to-one approach in the rather limited (especially in terms of network dynamics) data has been based on the assumptiosn that may not be valid - especially in the case of the crisis. It is expected that the customers, in the case when they have to shrink their orders, would be selective in the choice of the suppliers and they would not shrink their orders uniformly as our model assumes. However in order to know how the customers make these decision we would need to have more data.
  
  As a final note, the macroscopic effect that was captured by the
  measurements in figure \ref{fig:3growth}
  is realistic
  and its interpretation is supported by the official statistics
  on industrial production and business confidence in the
  given period: the statistics show that the Italian
  industrial production peaked in 2007 and then declined,
  reaching a 10-year low in 2009.

\section*{Acknowledgments}
  The work presented in this article is partly supported by the
  project ``A Large Scale Network Analysis of Firm Trade Credit'',
  project grant \texttt{IN01100017}, Institute for New Economic Thinking
  (INET).

\end{document}